\documentclass[manuscript,nonacm]{acmart}

\usepackage{booktabs}
\usepackage[htt]{hyphenat}
\usepackage{array}
\usepackage{enumitem}
\usepackage[acronym,nonumberlist]{glossaries}
\glsdisablehyper

\graphicspath{{./}}

\newacronym{AI}{AI}{artificial intelligence}
\newacronym{CAST}{CAST}{Context, Access, Stratum, and Timescale}
\newacronym{CT-VAR}{CT-VAR}{continuous-time vector autoregression}
\newacronym{DDP}{DDP}{data-donation package}
\newacronym{DSEM}{DSEM}{dynamic structural equation modeling}
\newacronym{ESM}{ESM}{experience sampling method}
\newacronym{EU}{EU}{European Union}
\newacronym{GDPR}{GDPR}{General Data Protection Regulation}
\newacronym{ID}{ID}{identifier}
\newacronym{LLM}{LLM}{large language model}
\newacronym{OS}{OS}{operating system}
\newacronym{VAR}{VAR}{vector autoregression}

\newglossaryentry{IOS}{name={iOS},description={Apple's mobile operating system}}
\newglossaryentry{AWARE}{name={AWARE},description={Mobile Context Instrumentation Framework}}

\renewcommand\footnotetextcopyrightpermission[1]{}

\title[A multilevel framework for social media measurement]{The CAST-framework: Measure and model social media use as a multi-level phenomenon through real-world applications}

\author{David Grüning}
\email{gruening@stanford.edu}
\affiliation{%
  \institution{Stanford University}
  \city{Palo Alto}
  \state{CA}
  \country{USA}}
\affiliation{%
  \institution{Max Planck Institute for Human
    Development}
  \city{Berlin}
  \country{Germany}}
\affiliation{%
  \institution{University of Cambridge}
  \city{Cambridge}
  \country{UK}}

\author{Jasper Doeninghaus}
\affiliation{%
  \institution{Myranda}
  \city{New York}
  \state{NY}
  \country{USA}}
\affiliation{%
  \institution{Harvard University}
  \city{Cambridge}
  \state{MA}
  \country{USA}}

\author{Zina Efchary}
\affiliation{%
  \institution{University of Cambridge}
  \city{Cambridge}
  \country{UK}}
\affiliation{%
  \institution{Myranda}
  \city{London}
  \country{UK}}

\author{Yui Kondo}
\affiliation{%
  \institution{MIT Media Lab, Massachusetts Institute of Technology}
  \city{Cambridge}
  \state{MA}
  \country{USA}}

\author{Kevin Dunnell}
\affiliation{%
  \institution{MIT Media Lab, Massachusetts Institute of Technology}
  \city{Cambridge}
  \state{MA}
  \country{USA}}

\author{Lennart Fischer}
\affiliation{%
  \institution{one sec}
  \city{Berlin}
  \country{Germany}}
\affiliation{%
  \institution{Technical University of Aachen}
  \city{Aachen}
  \country{Germany}}

\author{Isabella Zimmermann}
\affiliation{%
  \institution{Structured}
  \city{Berlin}
  \country{Germany}}

\author{Linnea Körte}
\affiliation{%
  \institution{Structured}
  \city{Berlin}
  \country{Germany}}

\author{Leonard Mehlig}
\affiliation{%
  \institution{Structured}
  \city{Berlin}
  \country{Germany}}

\author{Frederik Riedel}
\affiliation{%
  \institution{one sec}
  \city{Berlin}
  \country{Germany}}

\author{Paul Schmiedmayer}
\affiliation{%
  \institution{Stanford University}
  \city{Palo Alto}
  \state{CA}
  \country{USA}}

\authorsaddresses{Corresponding author: David Grüning,
  \href{mailto:gruening@stanford.edu}{gruening@stanford.edu}.}

\begin{abstract}
Designing social media experiences that support well-being requires understanding when, how, and for whom use matters.
Screen-time totals omit content and context, and connecting these with behavior and experience requires coordinating measurements across timescales.
We introduce the \gls{CAST} framework to connect measurement choices with person-specific models of exposure, behavior, physiology, and experience.
Its dimensions specify where observations occur, how they are obtained, what they measure, and at what temporal resolution.
Responses to interventions, such as whether to proceed after an app-opening pause, enter as behavioral measurements.
We propose four synchronized measurement modules linking mobile and wearable data with self-reports and intervention responses.
A synthetic demonstration with 120 simulated participants over 28 days illustrates how daily aggregation can obscure opposing effects of different activities under specified generating assumptions.
The framework guides selection of measures and outcomes for evaluating social media interfaces and interventions.
\end{abstract}



\begin{document}

\maketitle
\glsresetall

\section{Introduction}
\label{sec:intro}
Designing social media experiences that support well-being requires understanding when, how, and for whom use matters. The European Commission's 2025 guidelines for protecting minors recommend disabling features such as autoplay and push notifications by default~\cite{europeancommission2025minors}.
Evaluating these measures requires understanding which behaviors and experiences they change, for whom, and under what conditions.

For interventions targeting problematic smartphone use, fewer app openings alone cannot establish improvements in health or daily functioning.
Wearable sensors offer longitudinal estimates of sleep, physical activity, and heart rate~\cite{schmiedmayer2026design}.
Connecting these measures with mobile use logs and self-reports would let researchers examine whether changes in use accompany changes in health and perceived well-being.

Pairing an aggregate usage measure with a well-being score leaves four measurement problems unresolved. First, behavioral summaries depend on how use is measured.
A meta-analysis found that self-reported digital media use correlated only moderately with logged use~\cite{parry2021discrepancies}.
Treating either as a transparent measure of ``how much social media'' conflates substantive signals with method variance. Second, a duration summary leaves the activity and its context unspecified.
In a panel study linking Facebook logs and surveys, receiving targeted, composed communication from close ties was associated with improved well-being~\cite{burke2016relationship}.
However, a scoping review found limited support for the broader claim that active use benefits well-being while passive use harms it~\cite{valkenburg2022associations}.
Measures therefore need to identify the specific activity and social context rather than treating duration or an active/passive label as sufficient. Third, a measure at one timescale cannot represent every process of interest.
A daily total cannot resolve the immediate affective aftermath of a session.
Within-person changes also need to be distinguished from between-person associations.
Conventional cross-lagged panel models can misrepresent within-person relationships when they fail to separate stable differences between people from changes within them~\cite{hamaker2015critique}. Fourth, a pooled association can obscure differences between people.
An experience-sampling study found positive and negative associations between passive social media use and momentary well-being across participating adolescents~\cite{beyens2020effect}.
This variation motivates measurements that support person-specific estimates alongside population summaries.

These distinctions help connect interface design with behavior and experience.
\citet{lukoff2018meaningful} found that habitual smartphone use was associated with lower meaningfulness than use for a specific purpose.
In a survey, YouTube users generally described autoplay as reducing their sense of control, while search and playlists supported it~\cite{lukoff2021agency}.
Evaluating a design change therefore requires measures of the behavior and experience it is intended to change.

Existing tools already connect several measurement sources: \gls{AWARE} combines device sensing with mobile questionnaires~\cite{ferreira2015aware}.
Port supports local extraction and participant review of downloaded platform data before donation~\cite{boeschoten2023port}.
The question addressed here is how to coordinate such measurements with intervention responses and models that distinguish context, timescale, and person-specific relationships.

We propose to connect measurable layers of social media use with instruments that capture them at appropriate timescales.
Delivering interventions through those instruments would also let researchers study responses to a known prompt or friction alongside ongoing behavior.
The present paper consists of two parts.
The first develops the framework and its modeling implications; the second specifies a measurement architecture around four application modules and examines measurement choices in a synthetic demonstration.
The contribution links study-design decisions to the instruments, data, and models needed to evaluate social media experiences and interventions.

The aim is to connect a design choice, such as a change to a feed or an app-opening prompt, to the exposure, behavior, and experience that follow it.
The framework makes those measurement choices explicit while the proposed architecture identifies how they could be implemented together.

\section{Three senses of ``multilevel''}
\label{sec:senses}
We use multilevel in three distinct senses: stratification, nesting, and timescale.
Levels of stratification refer to the layers being studied: environment, behavior, physiology, and psychology.
Levels of nesting refer to the data structure: moments within sessions within days within persons within groups.
This is the statistical multilevel sense that motivates hierarchical models and within/between decomposition in general.
Levels of timescale refer to temporal resolution: a single interaction, a session, a day, a week, or a developmental span.
Different processes operate at different scales, and a measure fixed at one scale may miss processes at another.
The framework therefore addresses all three: stratification, nesting, and measurement and modeling across timescales.
Often existing designs are rich in one sense (e.g., dense nesting via experience sampling) while flat on another (a single stratum, a single timescale).

\section{The framework: a measurement design space}
\label{sec:framework}
\begin{figure}[t]
  \centering
  \includegraphics[width=\textwidth]{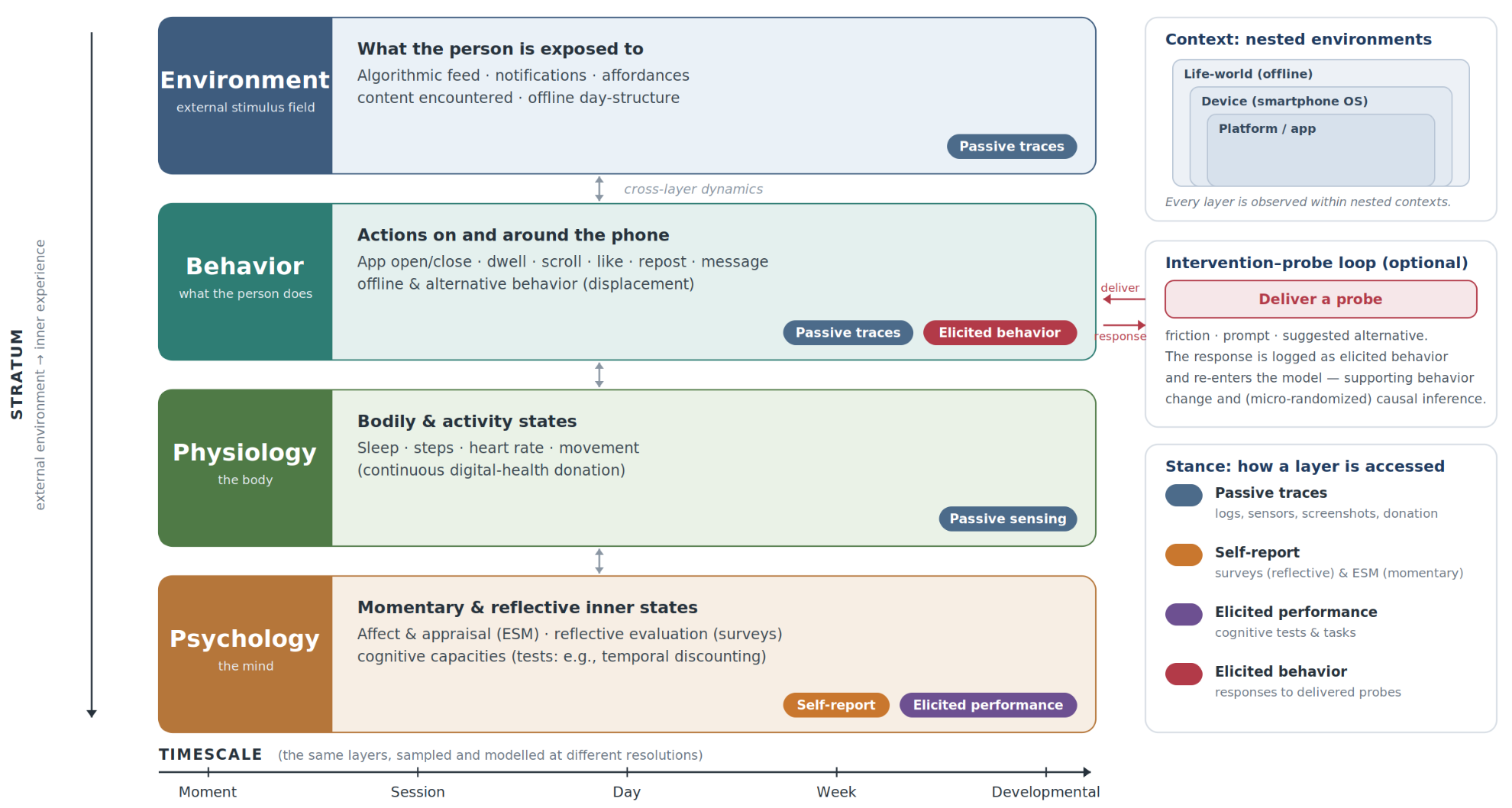}
  \Description{Diagram of the framework as a design space.
  Four stacked horizontal bands run down the figure, ordered from the external environment to inner experience.
  Environment, the external stimulus field: what the person is exposed to, including the algorithmic feed, notifications, affordances, content encountered and offline day-structure.
  Behavior, what the person does: app open and close, dwell, scroll, like, repost, message, and offline or alternative behavior as displacement.
  Physiology, the body: sleep, steps, heart rate and movement from continuous mobile and wearable sensing, contributed by health-data donation.
  Psychology, the mind: momentary and reflective inner states, including affect and appraisal by \glsentryshort{ESM}, reflective evaluation by survey, and cognitive capacities such as temporal discounting.
  Vertical double-headed arrows between adjacent bands mark cross-layer dynamics.
  Each band carries tags for stances that can access it, given as examples rather than as the only options, since behavior and context can also be self-reported; the tags are color-coded against a legend at the right titled Access, the measurement stance: passive traces (logs, sensors, screenshots, donation), self-report (surveys and \glsentryshort{ESM}), elicited performance (cognitive tests and tasks) and elicited behavior (responses to delivered probes).
  A timescale axis below the bands runs from moment through session, day and week to developmental, indicating that the same layers are sampled and modeled at different resolutions.
  A panel at the upper right shows three nested contexts: the offline life-world containing the smartphone device, which contains the platform or app.
  A second right-hand panel shows the optional intervention-probe loop, in which a probe (a friction, prompt or suggested alternative) is delivered to the behavior band and the logged response re-enters the model as elicited behavior.
  Randomizing the probe identifies the effect of the probe itself on subsequent use and experience; it does not identify the effect of use on well-being, which requires further assumptions.}
  \caption{Visualization of the multilevel framework for digital behavior and experience.
  Four strata (rows) accessed by different measurement stances (colors), across timescales and nested contexts, with cross-layer dynamics and an optional intervention-probe loop.
  Abbreviations: \glsentryshort{ESM}, \glsentrylong{ESM}; \glsentryshort{OS}, \glsentrylong{OS}.}
  \label{fig:framework}
\end{figure}

We propose a measurement design space with four dimensions.
We refer to this space as the \gls{CAST} framework.
Access denotes the measurement stance: how an observation is obtained.
Any measure of the social media use phenomenon can be located in this four-dimensional space.
For example, a sleep measure and a behavior log differ in stratum while both can be obtained through passive sensing.
Figure~\ref{fig:framework} shows the dimensions and an optional intervention loop whose responses enter the space as elicited behavior.
Evaluating an intervention requires specifying both its design and the measurements used to assess it.
\citet{mongeroffarello2026wellbeing} organize digital well-being by where an intervention exerts leverage, distinguishing self-oriented, collective, and systemic orientations of change, while \citet{lyngs2019selfcontrol} organize self-control tools by the psychological mechanism each design feature engages.
Both are frameworks of intervention.
\Gls{CAST} specifies what a study measures, by what means, at what temporal resolution, and in which context, making those choices explicit when evaluating interventions.

The four dimensions are not a taxonomy derived from first principles.
They are the axes along which existing measurement programs differ from one another, and along which their documented blind spots fall; Section~\ref{sec:contribution} locates those programs in the space and Table~\ref{tab:neighbors} lists what each misses.
Stratum follows the distinction that separates screen-content capture~\cite{reeves2020screenome} from sensor-based phenotyping~\cite{onnela2016phenotyping,torous2016tools} from survey and experience-sampling work, and the idiographic argument that structure at one level need not hold at another~\cite{molenaar2004manifesto}.
Access follows from the finding that logged and self-reported use diverge substantially and non-randomly~\cite{parry2021discrepancies}, which makes how an observation was obtained a property of the measurement rather than an implementation detail, and from the mobile experience-sampling literature that made the elicited stances routine~\cite{vanberkel2017esm}.
Timescale follows from continuous-time modeling, where the interval between observations changes the estimate itself~\cite{voelkle2012sem,driver2017ctsem,hamaker2018frontiers}.
Context follows both from the nesting of platform inside device inside offline life and from contextual integrity, where the appropriateness of a flow depends on the setting it occurs in~\cite{nissenbaum2004privacy,nissenbaum2010context}.
The fifth dimension of Section~\ref{sec:perturbation} was added because the intervention literature already delivers randomized stimuli~\cite{nahumshani2018jitai,klasnja2015mrt} whose logged responses are discarded as measurements once the effect is estimated.

\subsection{Stratum: the ``what'' (distance from the psychological layer)}
\label{sec:stratum}
We distinguish four strata, ordered roughly from the external stimulus field to the user psychology.
\textit{Environment} refers to what the person is exposed to: the algorithmic feed, notifications, affordances, the social field of a platform, and the structure of the offline day.
In social media much of this is partly exogenous and algorithmically generated.
Hence, treating ``content encountered'' as its own stratum, rather than folding it into behavior, is what lets us separate the exposure of content from a user's response to it.
\textit{Behavior} captures what the person does: app open/close, dwell and switch, scroll, like, repost, post, message, as well as alternative behaviors, offline (reading, exercise, sleep onset) or online (chess), that locate social media within a broader life system.
Alternative offline behavior can be self-logged or, with consent, inferred from passively collected location and movement data.
These data provide context for use episodes without additional reporting.
\textit{Physiology} concerns physiological and activity states: sleep timing and duration, step count, heart rate, movement.
In this framework, mobile and wearable sensor estimates complement self-reports of affect and perceived well-being.
Sleep and activity enter as possible antecedents or outcomes of use, rather than direct measures of subjective well-being.
\textit{Psychology} corresponds to the person's subjective experience and cognition: momentary affect and appraisal, reflective evaluations, and cognitive capacities and dispositions (e.g., temporal discounting, attentional control).
Treating exposure as its own stratum also makes it legible as a designed artifact rather than an ambient given.
Prior work has catalogued attention-capture designs in digital interfaces~\cite{mongeroffarello2023attention}.
Users report that interface features such as autoplay and recommendations can reduce their sense of control~\cite{lukoff2021agency}.
Measurement at this stratum is therefore measurement of an interface, which is what makes design changes testable within the same model as behavior and experience.

\subsection{Access: the ``how'' (measurement stance)}
\label{sec:stance}
Orthogonal to what is measured is how researchers gain access to it.
Four stances differ in burden, density, reliance on self-report, and whether the researcher controls the eliciting stimulus.
\textit{Passive traces} are accessible through device and in-app logs, screenshots, sensor streams, and data-donation packages.
Objective, continuous, low participant burden, but semantically thin (a like is logged; its meaning is not).
\textit{Elicited performance} is measured through cognitive tests and tasks.
Objective and semantically rich, but effortful and sparse; typically administered at a few points.
\textit{Self-report} captures subjective experience through surveys (reflective, retrospective, coarse-grained) and experience sampling (momentary, either time-triggered or event/activity-triggered).
Subjective and semantically rich, but reactive and burdensome.
\textit{Elicited behavior} is the result of deliberately introduced prompts, frictions, or nudges whose take-up, refusal, or latency is recorded.
An intervention probe provides a known stimulus whose behavioral response can be measured, just as a cognitive test elicits performance.
Section~\ref{sec:perturbation} describes how randomized delivery can support causal inference.
Mobile instrumentation frameworks support passive-trace collection~\cite{ferreira2015aware}, but those traces still require interpretation in relation to the construct being measured.
Mobile \gls{ESM} carries a substantial body of methodological guidance on trigger design, prompt scheduling, response rates, and participant burden~\cite{vanberkel2017esm}, and sensor-based work has begun to pair momentary affect with within-session signals during social media use specifically~\cite{ruensuk2020emotions}.
We treat that guidance as a binding design constraint on the protocol of Section~\ref{sec:protocol} rather than as an afterthought to it.

Separating stance from stratum distinguishes the physiological state being measured from passive sensing as the means of measurement.
Several strata can be accessed through passive observation or active elicitation.
The logged/self-reported divergence~\cite{parry2021discrepancies} is a difference between stances measuring the same stratum, making method variance part of the measurement problem.

\subsection{Timescale: resolution of metrics}
\label{sec:timescale}
Measures and processes are located on a temporal continuum: the moment (a single interaction and its immediate affective wake), the session, the day (rhythms, sleep--wake structure), the week (routines), and the developmental span (habit formation, well-being trajectories).
A guiding principle is scale-matching: a hypothesized process should be measured and modeled at the scale on which it operates, and cross-scale claims (a momentary process producing a developmental outcome) require an explicit account of aggregation.

\subsection{Context: nested environments}
\label{sec:context}
Behavior and experience occur within nested contexts: the offline life-world contains episodes of device use, which contain episodes within a specific platform or application.
The same behavior means different things in different contexts, and displacement (what an episode of use crowds out) is only visible when the surrounding context is measured.
Context is the environment stratum's structural counterpart and the reason Structured and alternative-behavior capture are not peripheral but load-bearing.

\subsection{The relational principle: meaning is configurational}
\label{sec:relational}
The central substantive commitment of the framework is that outcomes depend on configurations across cells of this space, not on any single cell.
The same behavior (thirty minutes on a platform) carries different significance depending on the environment cell (what content, from whom), the physiology cell (rested or sleep-deprived), the psychology cell (entered bored, left anxious), and the offline context (what it displaced).
The proposed model targets the joint distribution and cross-cell dynamics.
An association between one behavior measure and one outcome summarizes only part of that distribution and may obscure differences across configurations.

\subsection{Perturbation and response: intervention as eliciting behavior and impact}
\label{sec:perturbation}
Most measurement in this framework is observational: we watch the system and try to infer its dynamics.
But the same instruments that sense behavior can also act on it, and building a deliberate intervention layer into the framework serves two ends at once. Prior work has evaluated interventions for smartphone non-use~\cite{hiniker2016mytime} and app-opening frictions~\cite{gruening2023onesec,haliburton2024frictions}. Reviews organize these tools by their design mechanisms~\cite{lyngs2019selfcontrol} and identify limits in the evidence of effectiveness~\cite{mongeroffarello2023dswt}.
We place the response to a probe alongside other observations about the person, while retaining its role as an intervention outcome.
Two existing traditions already deliver stimuli and record what follows, and the difference is what they do with the record.
Just-in-time adaptive interventions and micro-randomized trials treat the response as an outcome, in order to estimate a proximal effect and to decide when delivery helps~\cite{nahumshani2018jitai,klasnja2015mrt}.
Control-theoretic and system-identification approaches treat the delivered input as a way to fit the dynamics of the system, so that the response informs a model of how the person's behavior evolves~\cite{rivera2007control,spruijtmetz2014dynamic}.
Both discard the response once its estimate is obtained.
We retain it as a measurement: a logged desistance is an observation about the person at that moment, in the behavior stratum, on the elicited stance, and it enters the measurement model of Section~\ref{sec:measmodel} alongside the passive and self-report indicators of the same construct.
The probe is in this sense a behavioral test item that can be administered repeatedly, in situ, at low burden, where a cognitive test can be administered only a few times, and it supplies the fourth stance that lets the psychology stratum be triangulated rather than measured one way.

The first is impact.
Much of this research is ultimately motivated by the wish to help people use these technologies better.
Delivering a brief friction, reflective prompt, or suggested alternative activity lets the same study measure ongoing use and evaluate an intervention.
The second is identification and measurement.
A delivered intervention provides a known stimulus, followed by a measurable response: whether the person closes the app or proceeds, how quickly, and with what subsequent affect and behavior.
The intervention--response episode may provide indicators of momentary states or traits, subject to validation of that interpretation.
Concretely, we have two constructs in mind and one response for each.
Whether a person desists at a given qualifying open is a candidate indicator of momentary self-regulatory capacity, a state that should vary within a person across a day and covary with sleep, time of day, and momentary affect.
That person's average desistance rate across many probes, and the latency of those responses, are candidate indicators of a more stable disposition, which should be comparatively stable across weeks and should relate to the temporal-discounting and attention tests administered at intake.
Validating either reading requires a study that is not the one that estimates the intervention's effect: convergence with \gls{ESM} appraisals at the moment level and with the cognitive tests at the person level, test--retest stability of the person-level component across the field period, and evidence that the component predicts later use beyond what passive traces already predict.
Absent that evidence, probe responses remain intervention outcomes with a plausible second reading, and we treat them as such.
Randomizing probe delivery adds an experimental component for estimating proximal effects of the probe, as discussed in the section on inference.

The response re-enters the design space as data in the behavior stratum, linked to the delivery that elicited it.
The framework preserves a probe-free base configuration for observational estimation and adds an active configuration for intervention studies.
Experience sampling and repeated testing can still introduce reactivity in the base configuration.

The probes discussed so far act on the behavior stratum.
A second class of interventions perturbs the environment stratum itself, by steering or re-ranking what a feed presents, muting a source, or surfacing alternative content, with the behavioral and experiential response observed downstream. Where platform access permits, these interventions manipulate presented content rather than prompting the person directly. The same logging and randomization requirements apply, so that steered exposure remains analytically separable from organic exposure.

\begin{table}[t]
  \centering
  \caption{The design space for capturing social media use as a stratified, multi-timescale phenomenon.
  Abbreviation: \glsentryshort{DDP}, \glsentrylong{DDP}.}
  \label{tab:designspace}
  \setlength{\tabcolsep}{3pt}
  \footnotesize
  \begin{tabular}{@{}>{\raggedright\arraybackslash}p{0.160\textwidth}>{\raggedright\arraybackslash}p{0.220\textwidth}>{\raggedright\arraybackslash}p{0.200\textwidth}>{\raggedright\arraybackslash}p{0.160\textwidth}>{\raggedright\arraybackslash}p{0.200\textwidth}@{}}
    \toprule
    \textbf{Suggested layer} & \textbf{Stratum} & \textbf{Stance} & \textbf{Typical timescale} & \textbf{Context} \\
    \midrule
    Smartphone-environment behavior & Behavior (+ Environment: notifications) & Passive traces & Moment--session & Device \\
    \addlinespace[2pt]
    Social-media-environment behavior & Behavior + Environment (content encountered) & Passive traces (logs, screenshots, \glsentryshort{DDP}) & Moment--session & Platform \\
    \addlinespace[2pt]
    Alternative behavior (offline/online) & Behavior & Passive traces + self-report & Session--day & Offline / other apps \\
    \addlinespace[2pt]
    Location and movement data & Behavior (offline activity) + Environment (offline setting) & Passive sensing & Continuous (moment--day) & Offline \\
    \addlinespace[2pt]
    Surveys & Psychology & Self-report (reflective) & Study-frame (pre/post) & Cross-context \\
    \addlinespace[2pt]
    Experience sampling & Psychology (+ momentary appraisal) & Self-report (momentary) & Moment--day & Time- or event-anchored \\
    \addlinespace[2pt]
    Tests & Psychology & Elicited performance & Study-frame & Cross-context \\
    \addlinespace[2pt]
    Physiological / digital health & Physiology & Passive sensing & Continuous (day--week) & Offline / device \\
    \bottomrule
  \end{tabular}
\end{table}

\section{From framework to models}
\label{sec:models}
The framework connects measurement choices to model specification.
Two model components are needed, and they should not be conflated: a measurement model that relates observed indicators to latent constructs across modalities, and a structural/dynamic model that relates the constructs to one another over time.

\subsection{Measurement model: multimodal indicators and method variance}
\label{sec:measmodel}
A single construct (e.g., ``engagement with a platform'') can be indexed by several stances (logs, screenshots, \gls{ESM} appraisal).
A measurement model with method factors, in the multitrait--multimethod tradition, separates trait variance shared across stances from method variance specific to a stance.
This reframes the logged/self-reported divergence not as a nuisance to be argued away but as an estimable quantity.
Measurement invariance should be examined both across persons and, in intensive data, across the within- and between-person levels, since a construct's indicators need not behave identically across levels.

\subsection{Dynamic structural model: cross-layer influence over time}
\label{sec:dynamic}
For within-person dynamics we favor multilevel \gls{VAR} / \gls{DSEM}~\cite{asparouhov2018dsem,hamaker2018frontiers}, which estimates contemporaneous and lagged relations among layers with person-specific random effects.
This represents individual dynamics and their distribution across the population.
Cross-layer, bidirectional effects are the object of interest: poor sleep (physiology) predicting next-day passive use (behavior) predicting evening affect (psychology) predicting sleep onset (physiology), closing a loop.
For streams sampled irregularly and at different rates, continuous-time formulations (e.g., ctsem~\cite{driver2017ctsem}; \gls{CT-VAR}~\cite{voelkle2012sem}) can represent elapsed time explicitly rather than treating unequal observation intervals as a fixed lag.

\subsection{Data integration across timescales: the asynchronous multi-rate problem}
\label{sec:datafusion}
A defining technical challenge is integrating streams sampled at different rates: screenshots or taps many times per minute, sleep and steps daily, and surveys and tests a few times per study.
Three strategies, used in combination: event-anchoring (align other streams to behaviorally defined events, e.g., the window around each platform session); aggregation to a shared grain (summarize fine streams to the coarsest scale of interest while preserving distributional features, not just means); and latent state-space models with mixed timescales (a slow latent state driving fast observations).
The framework's timescale dimension is what makes these choices explicit rather than ad hoc.

\subsection{Inference, heterogeneity, and person-specificity}
\label{sec:inference}
Causal claims from the observational data require explicit assumptions, ideally encoded as directed acyclic graphs, and separation of selection/state-dependence (people use social media in particular states) from effects (use changes states).
Within-person centering helps, but the strongest lever is the active layer described in Section~\ref{sec:perturbation}: randomized probe delivery.
Micro-randomized designs assign interventions at decision points, supporting estimation of proximal causal effects of the randomized intervention under the specified availability and analysis conditions~\cite{klasnja2015mrt}.
Randomized probe delivery does not by itself identify the causal effects of subsequent, nonrandomized social media use.
Random-effect distributions and mixture models can represent differences between people.
Specification-curve analysis addresses a separate question: how estimates change across analytical choices~\cite{orben2019association}.

\subsection{From questions to models}
\label{sec:questions}
To be usable, the framework must connect each type of scientific question to a concrete analysis.
Table~\ref{tab:questions} makes it explicit for the question types this program most often poses, naming for each the estimand, a suitable model class, and the assumptions identification rests on.
These examples connect each question to the data and assumptions needed to answer it.

\begin{table}[t]
  \centering
  \caption{Estimands, model classes, and identification assumptions for the program's core question types.
  Abbreviations: \glsentryshort{VAR}, \glsentrylong{VAR}; \glsentryshort{DSEM}, \glsentrylong{DSEM}.}
  \label{tab:questions}
  \setlength{\tabcolsep}{3pt}
  \footnotesize
  \begin{tabular}{@{}>{\raggedright\arraybackslash}p{0.200\textwidth}>{\raggedright\arraybackslash}p{0.260\textwidth}>{\raggedright\arraybackslash}p{0.250\textwidth}>{\raggedright\arraybackslash}p{0.230\textwidth}@{}}
    \toprule
    \textbf{Question} & \textbf{Estimand} & \textbf{Model class} & \textbf{Key identification assumptions} \\
    \midrule
    Does a probe change behavior in the moment? & Proximal within-person causal effect at decision points & Micro-randomized trial estimators~\cite{klasnja2015mrt} & Randomized delivery; correctly defined availability \\
    \addlinespace[2pt]
    How do layers influence one another over time? & Contemporaneous and lagged cross-layer effects within persons & Multilevel \glsentryshort{DSEM}; continuous-time models~\cite{asparouhov2018dsem,driver2017ctsem} & Explicit temporal assumptions; causal interpretation requires confounding control \\
    \addlinespace[2pt]
    Do different stances measure the same construct? & Trait versus method variance per construct & Multitrait-multimethod measurement model with method factors & Invariance across persons and levels \\
    \addlinespace[2pt]
    Who differs, and by how much? & Distribution of person-specific effects & Random-effect and mixture models & Sufficient within-person data; appropriate distributional assumptions \\
    \addlinespace[2pt]
    What do thin measures actually measure? & Calibration of coarse summaries against measured configurations & Decomposition or regression of survey totals on deep measurements & An overlap sample measured both ways \\
    \bottomrule
  \end{tabular}
\end{table}

\subsection{Reporting what was measured}
\label{sec:reporting}
The design space doubles as a reporting standard.
A study that locates itself in the space makes its measurement coverage and inferential limits explicit.
We therefore propose that studies of digital behavior and experience report, at minimum, the following ten items.

\begin{enumerate}[leftmargin=*,itemsep=1pt,topsep=3pt]
  \item The strata measured, and the strata explicitly left unmeasured.
  \item The stance used for each construct, and where constructs are triangulated across stances.
  \item The timescales sampled, with the scale-matching rationale for each hypothesis.
  \item The contexts covered (offline, device, platform) and whether displacement is observable.
  \item Whether each participant-period ran in the base or the active configuration.
  \item For active periods, the randomization scheme and the provenance logging of every delivery.
  \item The shared time base and the method used to align streams sampled at different rates.
  \item The measurement model, including how logged and self-reported divergence is handled.
  \item The within- and between-person decomposition, and how heterogeneity is reported.
  \item The missingness structure and the compliance model, specified in advance.
\end{enumerate}

\section{An integrated measurement system}
\label{sec:system}
\subsection{Design goals}
\label{sec:goals}
Our second contribution is a proposed modular architecture for implementing the framework.
Each module covers a region of the design space and is intended to write into a shared data model.
Existing tools and one new component provide the module implementations described below; the integrated system has not yet been deployed.
The goals: broad coverage across strata, stance triangulation on key constructs, dense-enough sampling on fast timescales, a shared identity and time base so streams can be fused, an optional active stance for intervention and probing, and a participant experience that is tolerable over weeks.
A further goal is portability.
The system is defined by its modules, its shared data model, and its interfaces rather than by any particular app, so the applications named below are best read as current implementations of an open specification that other tools can join, replace, or extend.

\subsection{Modules and their current implementation}
\label{sec:modules}
The system consists of four modules rather than a fixed set of applications (Table~\ref{tab:modules}).
A module is defined by the region of the design space it covers and the events it writes into the shared data model, not by the software that currently provides it.
The four modules cover the four strata across the nested contexts of Section~\ref{sec:context}.
Device behavior and assessment captures device-level use by passive traces, plus surveys, cognitive tests, and experience sampling.
Platform behavior and exposure captures within-platform behavior and content encountered through in-app capture or imported data-donation packages.
The donation route builds on local extraction from platform exports~\cite{boeschoten2022framework,boeschoten2023port}.
Offline context captures daily structure and displacement.
The physiology module is intended to link mobile and wearable records of sleep, activity, and heart rate with use episodes and self-reports.
The device module, and potentially the platform module, additionally hosts the optional decision-point engine of Section~\ref{sec:delivery}, which delivers interventions at decision points and records the response as elicited behavior.
Together the modules span all four strata, triangulate the psychology stratum across three stances, and pair continuous passive streams with sparse elicited and self-report measures.

Each module has one candidate implementation. one sec provides app-opening frictions and response logging~\cite{gruening2023onesec,haliburton2024frictions}.
The proposed architecture assigns one sec to the device module, Myranda to the platform module, Structured to the offline-context module, and a health-data donation component to the physiology module.
Status differs across modules and we mark it throughout: the instruments themselves are deployed applications, the health-data donation component is new, and the integration layer described in Section~\ref{sec:architecture} (shared identity, shared clock, shared schema) together with the platform-anchored experience sampling of Table~\ref{tab:modules} is specified here but not yet implemented.
Versioned releases of each implementation will be listed with the supplementary material.
These applications are examples, not requirements.
Any tool that writes the same schema-conformant events can implement a module, and an application can grow within its module over time, for instance by adding passive collection of location and movement data to offline context (Table~\ref{tab:designspace}) or platform-level decision points to the platform module.
The system requires that each module be covered, not that these tools cover it.

\begin{table}[t]
  \centering
  \caption{Modules of the measurement system mapped onto the design space, with the application that currently implements each.
  Abbreviation: \glsentryshort{ESM}, \glsentrylong{ESM}.}
  \label{tab:modules}
  \setlength{\tabcolsep}{3pt}
  \footnotesize
  \begin{tabular}{@{}>{\raggedright\arraybackslash}p{0.090\textwidth}>{\raggedright\arraybackslash}p{0.110\textwidth}>{\raggedright\arraybackslash}p{0.160\textwidth}>{\raggedright\arraybackslash}p{0.080\textwidth}>{\raggedright\arraybackslash}p{0.110\textwidth}>{\raggedright\arraybackslash}p{0.380\textwidth}@{}}
    \toprule
    \textbf{Module} & \textbf{Primary strata} & \textbf{Stance} & \textbf{Context} & \textbf{Timescale} & \textbf{Contribution to the shared model} \\
    \midrule
    one sec & Behavior, Psychology & Passive traces; self-report; elicited tests; (optional) elicited behavior & Device & Moment--session; study-frame & Device-environment behavior (app open/close, timing, duration) by passive sensing in the base configuration; baseline/endline surveys; cognitive tests; time- and event-triggered \glsentryshort{ESM}; and, in the optional active configuration, delivery of frictions/nudges as randomized probes whose responses are logged as elicited behavior \\
    \addlinespace[2pt]
    Myranda & Behavior, Environment, Psychology & Passive traces; self-report & Platform & Moment--session & Within-platform behavior (scroll, dwell, like, repost, message; content encountered), captured in an in-app platform browser (\gls{IOS}) and via imported data-donation packages, with all processing on-device; platform-anchored experience sampling (planned); candidate host for platform-level decision points (Section~\ref{sec:delivery}) \\
    \addlinespace[2pt]
    Structured & Behavior, Environment (offline) & Self-report / structured logging & Offline life-world & Day--week & Everyday structure, routines, and alternative (offline/online) behavior: the displacement and co-occurrence context \\
    \addlinespace[2pt]
    Health-data donation (new) & Physiology & Passive sensing (donation) & Offline / device & Continuous (day--week) & Native \gls{IOS} Health export: sleep, steps, activity, heart rate \\
    \bottomrule
  \end{tabular}
\end{table}

\subsection{Integration architecture}
\label{sec:architecture}
The system needs three shared foundations.
A common participant identity links a person's streams across apps and the donation package (via a study-issued \gls{ID} and enrollment flow).
A shared time base supports event-anchoring: timestamps must be reconciled to a common clock and time zone, with drift and offline buffering handled explicitly.
A unified data model defines events, sessions, prompts, responses, continuous samples, and, for the active layer, probe deliveries and their randomization draws.
Preserving instrument and stance provenance supports estimation of method factors and distinguishes probe-elicited from spontaneous behavior.
A minimum-necessary export principle should govern what enters the shared model.
Instruments perform feature extraction locally and contribute only derived, schema-conformant events (session boundaries, interaction events, content-category labels), while raw within-platform content remains on the device.
This keeps multi-instrument fusion compatible with on-device processing of the most sensitive streams and narrows the consent scope to what the analysis actually requires.
Consent, storage, and processing (ideally on-device or in a governed enclave for the most sensitive streams) sit atop this architecture, as discussed further in Section~\ref{sec:challenges}.
The fields the shared schema must carry are specified in Appendix~\ref{app:schema}.
We intend to release the full event schema, the reporting checklist of Section~\ref{sec:reporting} as a fillable file, and the synthetic demonstration corpus of Section~\ref{sec:demo} as open supplementary material, so the system is auditable and extensible beyond its current implementations.

This architecture also carries the program's privacy protections.
Contextual integrity requires attention to whether information flows respect the norms of their social context~\cite{nissenbaum2004privacy,nissenbaum2010context}.
Donation moves a platform footprint into a research context and consent legitimates that move without settling what may travel with it.
Combining streams can reveal information absent from individual records~\cite{solove2006taxonomy}.
Removing direct identifiers does not necessarily prevent re-identification~\cite{ohm2010broken}.
On-device processing limits what leaves the device, although derived events can still reveal sensitive information when linked.
In Myranda the feed is parsed in the in-app browser and only derived events are exported (session boundaries, dwell, a content-category label, and whether an item was recommended or followed); posts, authors, images and message text stay local.
Imported donation packages are parsed the same way and discarded after extraction, extending the local-extraction principle of the data donation literature~\cite{boeschoten2022framework} to continuous in-app capture.
The same boundary protects those never in a position to consent such as message partners and accounts appearing in a participant's feed, whose privacy is networked rather than individually held~\cite{marwick2014networked}.
Retaining raw content locally reduces the material exported for analysis; it does not eliminate the privacy risks of the fused record.
The cost is analytic: features not extracted on the device cannot be recovered later, so the shared schema and the reporting commitments of Section~\ref{sec:reporting} must be fixed in advance.

\begin{figure}[t]
  \centering
  \includegraphics[width=\textwidth]{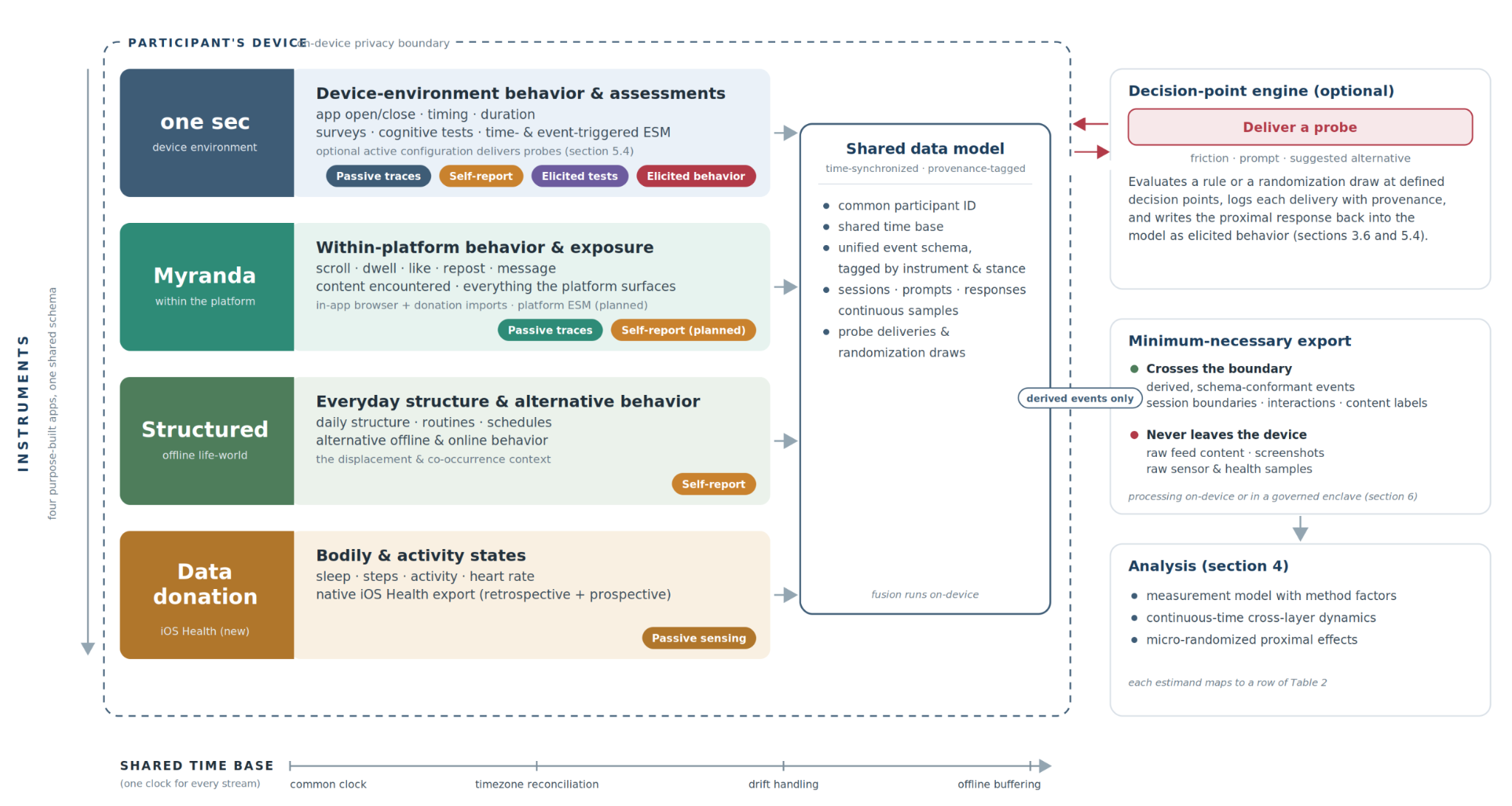}
  \Description{Diagram of the system architecture.
  At the left, four instruments sit inside a dashed boundary marking the participant's device.
  one sec covers the device environment: app open and close, timing and duration, plus surveys, cognitive tests and time- and event-triggered \glsentryshort{ESM}, and in its optional active configuration it delivers probes.
  Myranda covers within-platform behavior and exposure through an in-app browser and imported data-donation packages, contributing derived events rather than raw feed content: session boundaries, dwell, interaction events such as like, repost and message, content-category labels, and whether an item was recommended or followed.
  Structured covers the offline life-world: daily structure, routines, schedules and alternative offline and online behavior, giving the displacement and co-occurrence context.
  A health-data donation module covers bodily and activity states through mobile and wearable records exported from \gls{IOS} Health: sleep, steps, activity and heart rate, carried on the shared time base.
  Each instrument is tagged with the stances it supplies.
  Arrows carry their events into a shared data model in the center, which is time-synchronized and provenance-tagged and holds a common participant \glsentryshort{ID}, a shared time base, a unified event schema tagged by instrument and stance, sessions, prompts, responses, continuous samples, and probe deliveries with their randomization draws; fusion runs on-device.
  An optional decision-point engine reads and writes the same model, evaluating a rule or randomization draw at defined decision points, logging each delivery with provenance and writing the proximal response back as elicited behavior; it runs on the device inside whichever instrument hosts the decision point, and is drawn outside the boundary in the figure only for legibility.
  A minimum-necessary export gate labeled derived events only separates what crosses the boundary (derived, schema-conformant events: session boundaries, interactions, content labels, and summarized sleep, activity and heart-rate records) from what never leaves the device (raw feed content, screenshots, and raw sensor and health samples).
  Downstream, an analysis panel lists the measurement model with method factors, continuous-time cross-layer dynamics, and micro-randomized proximal effects.
  An axis along the bottom gives the shared time base: one clock for every stream, covering a common clock, timezone reconciliation, drift handling and offline buffering.}
  \caption{Proposed system architecture.
  Four instruments write provenance-tagged, schema-conformant events into a shared, time-synchronized data model, the on-device decision-point engine reads and writes the same model, and the on-device privacy boundary marks what never leaves the participant's device.
  The instruments are deployed applications; the shared identity, clock and schema that join them, the platform-anchored experience sampling of Table~\ref{tab:modules}, and the health-data donation module are specified here and not yet implemented.
  Abbreviations: \glsentryshort{ESM}, \glsentrylong{ESM}; \glsentryshort{ID}, \glsentrylong{ID}.}
  \label{fig:architecture}
\end{figure}
\subsection{Intervention and probe delivery}
\label{sec:delivery}
The optional intervention module is delivered through a decision-point engine, which any implementation that hosts a decision point can carry; at present one sec carries it at device-level decision points, and Myranda could carry it at platform-level ones.
Decision points include opening a target app, a scheduled time, or a sensed state.
At each decision point, the engine evaluates a delivery rule or randomization draw and delivers a probe (a friction, a prompt, or a suggested alternative).
It records delivery provenance (what, when, to whom, and the randomization draw) and proximal responses (compliance, latency, subsequent behavior and affect) in the shared model.
Base studies disable probe delivery; active studies enable it, optionally under micro-randomization.
Provenance-tagged delivery records distinguish probe episodes from surrounding behavior, subject to the reactivity and carryover considerations in Section~\ref{sec:challenges}.

\subsection{Illustrative protocol}
\label{sec:protocol}
The following three-to-four-week protocol illustrates the proposed integration.
At intake, one sec administers baseline surveys and cognitive tests (temporal discounting, attention).
Across the field period, one sec passively logs device-level behavior and delivers \gls{ESM} prompts (a mix of time-triggered and event-triggered, e.g., on opening a target app); Myranda logs within-platform behavior and content encountered and delivers platform-anchored \gls{ESM}; Structured captures daily structure and alternative behavior; and the donation component collects continuous \gls{IOS} Health data (retrospective at enrollment plus prospective through the field period).
At exit, one sec repeats surveys and tests.

In an active variant, one sec's decision-point engine micro-randomizes a friction on opening a target app: at each qualifying open, a probe is delivered with some probability, its delivery logged, and the proximal response (proceed vs. desist, latency, next app, subsequent \gls{ESM} affect) recorded as elicited behavior.
As outlined in Section~\ref{sec:models}, analysis then instantiates: a measurement model with method factors on the triangulated constructs; a continuous-time model for cross-layer dynamics; event-anchored fusion of the slower streams to platform sessions; proximal causal effects estimated from the micro-randomized design; and a specification-curve/heterogeneity layer over person-specific estimates.
Both arms should be pre-registered, with template consent language for the active layer released as supplementary material.
The active arm requires a power calculation that accounts for the repeated randomization and participant availability~\cite{klasnja2015mrt}.
That calculation is specified by four quantities, which a study should fix and report in advance: the proximal outcome (desistance at a qualifying app open, a binary decision-point outcome); the smallest proximal effect the study aims to detect, on the log-odds scale; the expected number of available decision points per participant, as qualifying opens per day multiplied by field days and by the availability rate; and the randomization probability at each available decision point.

\subsection{A worked demonstration}
\label{sec:demo}
For demonstration, we present a fully synthetic demonstration that examines how measurement and modeling choices affect recovery of known effects under a specified generating process.  The materials for replicating this simulation can be accessed in the manuscript's OSF-repository: \url{https://osf.io/kvaw9/}. Appendix~\ref{app:reproduction} describes all files needed for the simulation in detail.

A panel of 120 simulated participants over 28 days, matching the illustrative protocol, is generated across all four streams.
The generating model specifies short sleep raising next-day passive dwell, whose algorithmic content share depresses concurrent affect through a kernel with a 45-minute half-life.
Evening affect feeds back onto sleep, and dwell inside the person's pre-sleep block additionally delays sleep.
The boundary of that block is recorded only in the Structured stream.
Half the panel runs probe-free, as described in Section~\ref{sec:perturbation}; the active arm adds randomized probes.
The two arms are analyzed separately and never pooled: the design ladder, the continuous-time model, and the context and measurement analyses use the base arm, so that no estimate is contaminated by probe-elicited behavior, while the proximal probe effect and the instrumental analysis use the active arm.
The streams are fused into the shared data model and analyzed under a ladder of five designs sharing one day-level estimand, each adding a measurement or modeling choice.
The daily and continuous-time analyses do not target the same quantity, and their recovery percentages are not directly comparable.
The day-level estimand is the total effect of a day's exposure on that evening's affect, in affect points per daily exposure minute.
The continuous-time estimand is the instantaneous effect of exposure on concurrent affect, in affect points per kernel-weighted minute under a 45-minute half-life; the day-level quantity is an aggregation of it over a day, and each estimate is reported as a share of its own target.
The comparison examines recovery of the planted effects across these designs and recovery of a planted proximal effect from a micro-randomized probe sequence.
In the generating model, short sleep raises next-day passive dwell by 14.3 minutes per hour of sleep lost.
Algorithmic exposure lowers concurrent affect by 0.122 scale points per kernel-weighted minute while close-tie exchange raises it by 0.147.
Evening affect feeds back onto sleep at 0.23 hours per unit, and pre-sleep dwell displaces sleep at 0.55 minutes per minute.

\begin{figure}[t]
  \centering
  \includegraphics[width=\textwidth]{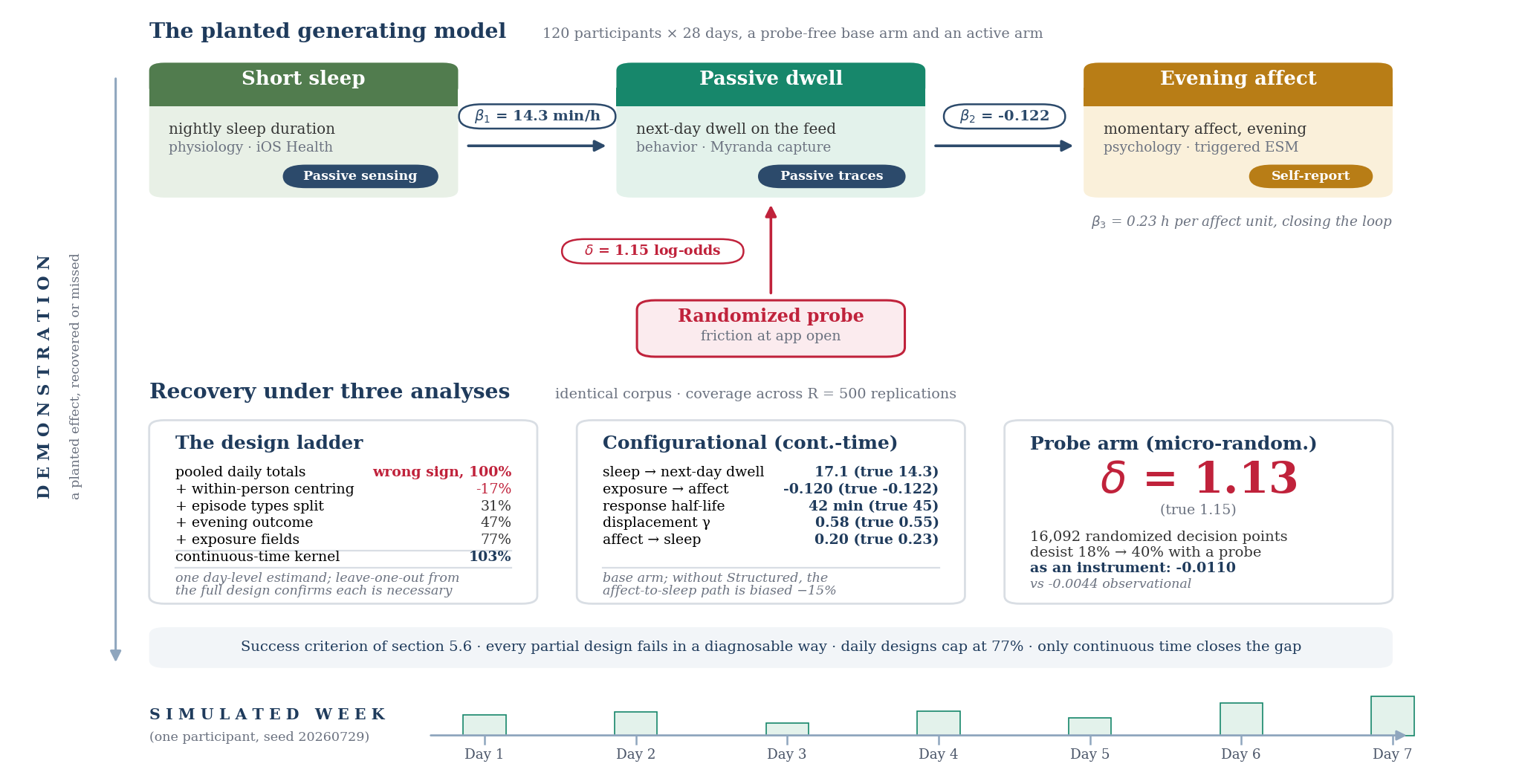}
  \Description{Three-part figure showing a planted effect in fully simulated data and whether each analysis recovers it.
  At the top, the generating model appears as three linked boxes: short sleep (nightly sleep duration, physiology stratum, from \gls{IOS} Health) feeds passive dwell (next-day dwell on the feed, behavior stratum, from Myranda) with a coefficient of 14.3 minutes per hour; passive dwell feeds evening affect (psychology stratum, from triggered \glsentryshort{ESM}) with a coefficient of -0.122; and a path of 0.23 hours per affect unit closes the loop back to sleep.
  A randomized probe, a friction at app open, raises the log-odds that the person desists at that decision point by 1.15, which in turn reduces dwell.
  In the middle, three panels compare recovery from an identical corpus.
  The design ladder panel shows recovery as design elements are added: pooled daily totals give the wrong sign in 100 percent of replications, within-person centering reaches -17 percent, splitting episode types 31 percent, adding an evening outcome 47 percent, adding exposure fields 77 percent, and a continuous-time kernel 103 percent.
  The configurational continuous-time panel recovers sleep to next-day dwell at 17.1 against a true 14.3, exposure to affect at -0.120 against -0.122, a response half-life of 42 minutes against 45, displacement at 0.58 against 0.55, and affect to sleep at 0.20 against 0.23.
  The micro-randomized probe-arm panel recovers a probe effect of 1.13 against a true 1.15, from 16,092 randomized decision points, with desistance rising from 18 to 40 percent under a probe and an instrumental estimate of -0.0110 against -0.0044 observationally.
  A banner beneath states that every partial design fails in a diagnosable way, that daily designs cap at 77 percent, and that only continuous time closes the gap; these statements describe the designs and the generating model tested here and are not general claims.
  At the bottom, an unlabeled bar strip shows a simulated week for one participant, days 1 through 7.}
  \caption{Synthetic demonstration: the planted generating model and its recovery under the design ladder, the continuous-time configurational model, and the micro-randomized probe arm.
  Simulating 120 participants for 28 days with a probe-free base arm and an active arm.
  All app, sleep and health records are simulated; no participant data were used.
  The ladder and the leave-one-out comparison rest on 24 draws, the continuous-time and probe analyses on 500 replications, and intervals are reported at the 95 percent level.
  Recovery (the share of a target effect estimated) and interval coverage are distinct quantities and the daily and continuous-time panels target different estimands, so their percentages are not directly comparable.
  Statements in the figure describe the designs and effects tested in this generating model.
  Abbreviation: \glsentryshort{ESM}, \glsentrylong{ESM}.}
  \label{fig:demo}
\end{figure}

Recovery describes the share of a target effect estimated; interval coverage describes how often an interval contains that target across replications.
Both are reported descriptively: no threshold separates success from failure, and none was fixed in advance, so the ladder should be read as an ordering of designs rather than as a set of passes and failures.
Intervals are reported at the 95 percent level.
The analyses also differ in how many replications support them, and the counts are not interchangeable: the ladder rungs and the leave-one-out comparison rest on 24 draws, while the continuous-time and probe analyses rest on 500 replications.
The ladder reads as follows, each rung a share of the same day-level exposure path.
Pooled daily totals return the wrong sign in every replication ($p < 10^{-5}$), and within-person centering recovers $-$17 percent.
In this generating model, the daily total combines episode types with opposing effects and exposure occurring at different times relative to the affect measurement.
Separating episode types recovers 31 percent, matching the outcome to the evening 47 percent, and adding Myranda's exposure fields, the environment stratum of Section~\ref{sec:stratum}, 77 percent.
This daily design still leaves roughly a quarter of the specified path unrecovered.
The continuous-time model then targets the finer-scale exposure path.
Under the continuous-time kernel, its half-life profiled rather than assumed, the model targets the finer path directly and recovers it at 103 percent, with coverage of 100 percent across the 24 ladder draws, which at that number of draws is consistent with true coverage as low as 88 percent.
The sleep-to-dwell path shows $-$3 percent relative bias and 92 percent coverage across the 500 replications.
The half-life is profiled on a grid that excludes the truth: its median estimate is 42 minutes against a true 45, the nearest available rung, and it lands within 20 percent of the truth in 99.8 percent of those replications.
The continuous-time estimate targets a finer quantity than the daily estimates and must be interpreted against that target.
In the active arm, 16,092 randomized decision points yield a proximal probe effect of $+$1.13 log-odds against a true $+$1.15, on the odds that the person desists at the decision point rather than proceeding into the session; the effect on dwell follows from desistance rather than being estimated directly.
These comparisons illustrate losses in recovery when measurements or temporal information used by the generating model are omitted.
Figure~\ref{fig:demo} summarizes the planted model and all three analyses.

A complementary comparison removes each design choice from the fully specified daily design to examine sensitivity to the ladder's order of addition.
Across the 24 ladder draws that design recovers 77 percent of the day-level path; dropping within-person centering collapses recovery to $-$3 percent, dropping episode separation to $-$25 percent, dropping timescale matching to 50 percent, and dropping exposure fields to 47 percent.
Each omission reduces recovery in the tested generating model.

Three further analyses examine context, measurement error, and randomized delivery.
Context: displacement is recovered at 0.58 against a true 0.55.
Without the Structured stream, a fixed bedtime is assumed and displacement is misattributed to the affect-to-sleep path, yielding $-$15 percent bias.
Measurement: even with context, that path uses an errors-in-variables correction whose reliability term is set by assumption in the simulation and would require calibration in a field study (Section~\ref{sec:conclusion}).
Coverage remains 82 percent under the simulated missing-not-at-random evening prompts.
The active-layer simulation assumes that randomized delivery shifts dwell only through desistance.
The daily count of delivered probes is then used as an instrument for that day's exposure, targeting the day-level effect of exposure on evening affect.
Three conditions make it a valid instrument in the simulation, and each is a condition a field study would have to defend rather than assume.
Relevance holds by construction: a delivered probe raises the odds of desisting, so days with more deliveries carry less exposure.
Independence holds by design: eligible decision points are qualifying opens of a target app, delivery at each is drawn with a fixed randomization probability, and the draw is therefore independent of the person, the day, and anything they share.
Exclusion is the substantive assumption: the friction must reach evening affect only by changing exposure, which fails if seeing a prompt itself alters mood or if it displaces the person into some other affect-relevant activity.
Because participants differ in how many qualifying opens they have, the number of chances to be randomized is itself a person characteristic; the instrument is therefore person-mean-centered and the count of available decision points enters as an offset, so that the estimate rests on within-person variation in draws rather than between-person variation in opportunity~\cite{klasnja2015mrt}.
This yields $-$0.0110 against $-$0.0044 observationally, with intervals overlapping in 98 percent of replications.
Interval overlap alone does not establish that the two estimates identify the same effect.

We hasten to note one limitation here in that the sleep-to-dwell path was planted at six minutes per hour to begin with and was not reliably detectable against roughly twenty-nine minutes per day of within-person variability.
We increased the coefficient during exploratory tuning until detection became reliable near 14.3 minutes per hour under the simulated conditions.
This tuning was exploratory and was not pre-specified: the resulting coefficient is the value at which this design can recover the path in simulation, not an estimate of a plausible field effect, and it should not be read as one.

\section{Challenges and considerations}
\label{sec:challenges}
Using the framework in a field study requires resolving privacy, participant burden, missingness, and integration constraints.

Screenshots, within-platform content, and health donations are among the most sensitive data a study can collect.
The program requires informed consent, data minimization, and, wherever possible, on-device processing or governed enclaves, with special attention to third parties captured incidentally (message partners, people in others' feeds).
If the study population includes minors, the consent architecture additionally requires parental consent alongside participant assent, age-appropriate disclosure, and safeguarding procedures.
Studies should also name the data controller and processor for the study-issued identity and the fused dataset, since accountability for the linked streams cannot be left implicit. For the present submission we note that no data were collected from human participants: the demonstration corpus of the demo is fully synthetic and generated by the authors. The proposed field protocol presented for illustration has not yet undergone institutional ethics review.

The active layer deliberately tries to change behavior, which raises the ethical bar beyond that of passive measurement. Consent must specifically cover intervention; randomization needs genuine equipoise; probes must avoid foreseeable harm such as frustration or the withholding of needed access; participants need control (opt-out, override, and a clear account of when the system will act); and researchers must weigh whether an effective intervention should be withheld from a control condition. Probe responses are themselves behavioral, sensitive data.

The base configuration disables intervention probes.
Experience sampling and repeated testing still induce some reactivity, and the active layer perturbs deliberately.
Studies must identify the configuration for each participant-period and account for possible reactivity and carryover (Sections~\ref{sec:perturbation} and~\ref{sec:inference}).

Multi-app protocols multiply the ways data can be missing, and missingness is rarely at random (people skip prompts in particular states).
Compliance, non-response models, and planned-missingness designs belong in the analysis plan from the start.

Four data sources, possibly different clocks and buffering behaviors, and offline periods make the shared time base (Section~\ref{sec:architecture}) a first-order engineering risk, since event-anchored fusion degrades quickly under misalignment.

Sensor-derived health measures require validation for the intended use~\cite{dunn2018wearables}.
Studies should document source devices, measurement coverage, and missing data alongside the donated values.

The health-donation component as described is \gls{IOS}-native; an \gls{IOS}-only sample constrains generalizability and should be acknowledged, with Android parity noted as future work. Platform coverage inside Myranda similarly bounds which social media environments the framework can actually observe.

Within-platform capture also depends on the platforms themselves. Export formats change, rate limits shift, and in-app browser capture can be affected by interface redesigns or terms-of-service enforcement. The program should treat this as a design constraint, grounding donation-based access in data-portability rights (\gls{GDPR} Article 20) and, in the \gls{EU}, the transparency and researcher-access provisions of the Digital Services Act, and version-controlling parsers so that platform changes degrade coverage rather than validity.

The system trades scalability for depth.
We position it not as a replacement for large, thin studies but as the intensive, mechanism-focused complement that tells us what the thin measures are actually measuring.

The integrated system's field feasibility remains untested.
The system described in Part 2 has not yet been deployed as an integrated whole, so its feasibility claims, namely tolerable burden across three to four weeks, workable compliance across four concurrent instruments, and comprehensible consent for a fused record, are arguments rather than findings.
The synthetic demonstration examines recovery under specified generating assumptions; it establishes neither the general necessity of every measurement nor participant acceptance of the integrated protocol.
Reporting compliance, dropout, and participants' understanding of the on-device boundary is therefore the first obligation of the calibration study in Section~\ref{sec:conclusion}, and we regard those numbers as a precondition for the rest of the program rather than as a byproduct of it.

\section{Contribution and positioning}
\label{sec:contribution}
The framework synthesizes several existing programs rather than competing with them.
It shares the fine-grained, in-situ measurement goals of the screenome program~\cite{reeves2020screenome} and digital phenotyping~\cite{onnela2016phenotyping,torous2016tools}.
It builds on donation-based access to platform data~\cite{boeschoten2022framework,boeschoten2023port}, including work using \glspl{LLM} to support reflection through hypothetical inferences over donated histories~\cite{kondo2025mirror}.
It also addresses the measurement differences documented in comparisons of logged and self-reported use~\cite{parry2021discrepancies}.
Specification-curve analysis informs the assessment of sensitivity to analytical choices~\cite{orben2019association}.
Its active configuration connects to just-in-time adaptive interventions~\cite{nahumshani2018jitai}, micro-randomized trials~\cite{klasnja2015mrt}, control-system approaches~\cite{rivera2007control} and dynamic models of behavior~\cite{spruijtmetz2014dynamic}, and single-case/N-of-1 experimental designs.

Table~\ref{tab:neighbors} locates these neighboring programs, together with a cross-sectional survey design, in the design space.
The comparison helps researchers identify which measurements a study includes and which its research question may also require.

Two of these programs are close enough to this proposal that the difference should be stated precisely rather than by category.
\Gls{AWARE} is a mobile instrumentation framework that collects device sensors and logs and administers experience sampling on the same device~\cite{ferreira2015aware}.
It covers the device context thoroughly and can carry the psychology stratum by self-report, but what happens inside a platform is not visible to it: an app-usage log records that a feed was open, not what it presented.
It also has no notion of a stance as a property of a measurement, and no place to record a randomization draw, so probe-elicited behavior cannot be distinguished from spontaneous behavior in the record it produces.
Port supports local extraction from platform data-donation packages, which does reach platform content~\cite{boeschoten2023port,boeschoten2022framework}.
That record is retrospective and event-stamped rather than continuous, carries no momentary experience, and includes neither device-level nor physiological streams, so exposure can be reconstructed while the state the person was in cannot.
What the integration adds is not a new sensor but the join: one participant identity, one reconciled clock, and one event schema that carries instrument and stance provenance and randomization draws across all four strata.
That join is what makes method factors estimable across stances, exposure alignable to affect within the minute rather than the day, and probe responses analyzable as measurements rather than only as outcomes.
Neither framework is displaced by this: both write events that a module could contribute, and \gls{AWARE} in particular is a plausible implementation of the device module for a team that already runs it.

\begin{table}[t]
  \centering
  \caption{Neighboring measurement programs located in the design space.
  Coverage varies by implementation; the rows describe the cited examples and their core measurements.
  Abbreviation: \glsentryshort{DDP}, \glsentrylong{DDP}.}
  \label{tab:neighbors}
  \setlength{\tabcolsep}{3pt}
  \footnotesize
  \begin{tabular}{@{}>{\raggedright\arraybackslash}p{0.130\textwidth}>{\raggedright\arraybackslash}p{0.150\textwidth}>{\raggedright\arraybackslash}p{0.140\textwidth}>{\raggedright\arraybackslash}p{0.120\textwidth}>{\raggedright\arraybackslash}p{0.130\textwidth}>{\raggedright\arraybackslash}p{0.260\textwidth}@{}}
    \toprule
    \textbf{Design} & \textbf{Strata} & \textbf{Stance} & \textbf{Timescale} & \textbf{Context} & \textbf{Blind spots} \\
    \midrule
    Cross-sectional survey (e.g.,~\cite{przybylski2017goldilocks}) & Psychology; behavior by recall & Self-report (reflective) & Study-frame & Weekday / weekend & Within-person dynamics and observed exposure absent; shared self-report method \\
    \addlinespace[2pt]
    Experience sampling~\cite{vanberkel2017esm} & Psychology; reported behavior and context & Self-report (momentary) & Moment to day & Prompt-dependent & Direct observation requires complementary sensing \\
    \addlinespace[2pt]
    Screenome~\cite{reeves2020screenome} & Behavior; environment (screen content) & Passive traces (screenshots) & Moment to session & Device & Sensitive content; physiology and subjective experience require complementary measures \\
    \addlinespace[2pt]
    Digital phenotyping~\cite{torous2016tools} & Behavior; psychology & Passive traces; self-report & Continuous; prompted & Device; offline & Coverage depends on sensors and surveys; platform content needs additional capture \\
    \addlinespace[2pt]
    Data donation~\cite{boeschoten2022framework} & Behavior; environment (platform) & Passive traces (\glsentryshort{DDP}) & Moment-stamped, retrospective & Platform & Platform-defined exports; momentary self-reports require additional collection \\
    \addlinespace[2pt]
    Integrated system (Part 2) & All four & All four, including elicited behavior & Moment to study-frame & Offline, device, platform & Scale, burden, \gls{IOS}-only sample \\
    \bottomrule
  \end{tabular}
\end{table}

The contribution has four parts: (a) a design space separating context, access, stratum, and timescale; (b) modeling requirements for relations across layers and timescales; (c) intervention responses treated as elicited behavior alongside passive traces and self-report; and (d) a proposed architecture linking these measurements across four modules.

Three consequences follow for how work in this area is built and judged.
The first is a design implication for the tools themselves. Friction tools already record app-opening attempts and responses~\cite{gruening2023onesec,haliburton2024frictions}.
Linking these records with context, experience, and randomization draws where applicable would support joint analysis of intervention responses and ongoing use.
This use requires an explicit logging design and participant consent, as described in Section~\ref{sec:challenges}.
The proposed event schema is intended to support that design.
The second concerns evaluation.
The synthetic demonstration illustrates how an outcome measure can obscure an effect under specified assumptions.
In that generating model, daily totals combine episode types with opposing effects and discard the timing of exposure relative to affect.
The implication for evaluation is to match outcomes to the behavior or experience an intervention is intended to change.
Linking mobile use records with wearable measures and self-reports would support evaluation of sleep, activity, and perceived well-being alongside changes in app use.
Whether measurement choices explain part of the heterogeneous effects reported for digital self-control tools~\cite{mongeroffarello2023dswt} remains an empirical question.
We do not propose a reanalysis of existing trials to settle it, because we know of no public dataset that carries episode type, exposure timing, and outcome timestamps together, which is the same gap the framework is meant to close.
Answering it requires a study instrumented for it in advance.
The third is about reporting.
The ten items of Section~\ref{sec:reporting} make the measurement choices and assumptions needed to interpret and compare studies explicit.
The on-device extraction boundary of Section~\ref{sec:architecture} offers a reusable design for limiting raw-content export across implementations.

\section{Conclusion and roadmap}
\label{sec:conclusion}
Studying social media use in context requires connecting exposure, behavior, physiology, and experience at the timescales relevant to the research question.
We propose a four-dimensional measurement framework, a strategy for modeling cross-layer dynamics within and between persons, and an architecture that incorporates intervention responses alongside other measurements.
Linking mobile use logs with wearable measures of sleep, activity, and heart rate would complement self-reports with objective indicators of behavior and physiology.
The synthetic demonstration examines these choices under specified generating assumptions; feasibility of the integrated system remains to be established in the field.

The next step is a three-study program.
A calibration study runs the full system alongside conventional measures in a modest sample, quantifying what screen-time totals and retrospective surveys actually measure, which bears directly on how the existing evidence base should be read.
An intensive observational study then estimates person-specific cross-layer dynamics in the base configuration.
An active study finally adds micro-randomized probes on a focused question.
Its evaluation should assess whether changes in app use accompany changes in sleep, activity, and reported well-being.
This sequence would move from calibration and feasibility to observational estimates and then to intervention evaluation.
The present contribution gives researchers a common basis for choosing what to measure, how to connect the measurements, and which claims those measurements can support.

\begin{acks}
DG's work is funded by Huo Family Foundation and Stanford's Center for Digital Health. DG has ongoing research projects in collaboration with the the apps.
PS's work is funded by Stanford's Center for Digital Health.
\end{acks}

\section*{Competing interests}
LF and FR develop one sec, IZ, LK, and LM develop Structured, and JD and ZE developed Myranda, three of the four instruments described in Part 2.
Yk is a Product Advisor and KD a Technical Advisor to Myranda.

\section*{Use of generative \glsentryshort{AI}}
Generative \gls{AI} assistants (Anthropic's Claude and OpenAI's Codex) were used to draft and revise portions of this manuscript and the figure descriptions.
They were also used to check the manuscript for internal consistency.
All study design, data collection, and analysis were carried out by the authors without generative \gls{AI}.
The authors reviewed and verified all content, including all reported statistics, and take full responsibility for the manuscript.



\appendix
\section{Fields of the shared event schema}
\label{app:schema}
The architecture of Section~\ref{sec:architecture} requires that every instrument write the same event type, whatever the instrument.
This appendix states the fields that event must carry for the analyses of Section~\ref{sec:models} to be possible; the schema itself, with types and validation rules, will be released as supplementary material.

Every event carries a study-issued participant identifier, an event identifier, and the identifier of the session it belongs to where one applies.
Every event carries a timestamp reconciled to a common clock, the originating time zone, and a flag recording whether the event was buffered offline and written later, since drift and buffering otherwise reach the analysis as spurious timing.
Every event carries two provenance fields that the measurement model of Section~\ref{sec:measmodel} requires and that cannot be reconstructed after the fact: the instrument that produced it, and the stance under which it was obtained, in the sense of Section~\ref{sec:stance}.
Events in the active layer carry three further fields: the decision point at which delivery was considered, whether the participant was available at it, and the randomization draw with the probability under which it was made.
Without the draw and the availability flag, probe episodes cannot be distinguished from spontaneous behavior and the proximal analyses of Section~\ref{sec:inference} are not identified.

Content-bearing events carry only derived fields, as required by the export rule of Section~\ref{sec:architecture}: session boundaries, dwell, interaction type, a content-category label, and whether an item was recommended or followed.
Raw feed content, screenshots, and raw sensor samples are not schema fields and do not leave the device.

\section{Reproduction materials for the synthetic demonstration}
\label{app:reproduction}
The demonstration of Section~\ref{sec:demo} is fully simulated. The release is provided as supplementary material with this submission in the manuscript's OSF-repository: \url{https://osf.io/kvaw9/}. It contains the generating model
(\texttt{generator/generate.py}) and its parameter file
(\texttt{generator/config.py}, which holds every planted path and coefficient
and every seed), the shared event schema, the emitted streams
(\texttt{corpus/streams/}), the analysis scripts for the design ladder
(\texttt{analysis/ablation.py}), the continuous-time model and the probe arm
(\texttt{analysis/analyses.py}), the replication harness
(\texttt{replication/run.py}), the self-report against the reporting
checklist of Section~\ref{sec:reporting} (\texttt{CHECKLIST\_4.6.md}), and
the saved outputs behind every value in Section~\ref{sec:demo} and
Figure~\ref{fig:demo}: \texttt{corpus/replication\_draws.csv} (500 draws),
\texttt{corpus/ladder\_draws.csv} (24 draws),
\texttt{corpus/replication\_performance.csv},
\texttt{corpus/headline\_numbers.json}, and
\texttt{figures/figure3\_manuscript.png}. \texttt{RELEASE.md} maps each
number in the text and the figure to the key and file it is read from.

Three seed ranges are used. The representative corpus, from which every point
estimate in the text and in Figure~\ref{fig:demo} is taken, is generated
from seed 20260729; the 24 ladder draws use seeds 30000 to 30023; the 500
replication draws use seeds 90001 to 90500. Person-level parameters are
redrawn under each seed, so coverage is scored against a moving target.

With the pinned environment installed (\texttt{requirements.txt}; CPython
3.14.2, statsmodels 0.14.6), \texttt{make reproduce} regenerates the streams,
both sets of draws, the producer file and the figure, and \texttt{make
verify} asserts that every number quoted in the accompanying documentation is
one the producer emits and that stored draws recompute exactly. On the
platform the release records (Apple M4 Pro, macOS 26.5, Accelerate BLAS) the
regeneration is bit-identical and every gate passes. The emitted corpus, every
point estimate, the ladder shares and every number in this section and in
Figure~\ref{fig:demo} also reproduce, at their reported rounding, on the two
other platforms checked (a 16-core Mac of an earlier generation and Linux
x86-64 with OpenBLAS). What varies across platforms is the mixed-model
interval bounds, by about two per cent of the interval half-width, and
therefore the coverage of the sleep-to-dwell, affect-to-sleep and
displacement paths, which moved by at most 1.2 percentage points, within
Monte Carlo error. \texttt{RELEASE.md} records the comparison.

\end{document}